**Electromotive force generation in a ferromagnetic metal thin film under the ferromagnetic resonance excitation with permanent magnets**

Ryutoku Fujii,[1] and Eiji Shikoh [1,a)]

AFFILIATIONS

[1]*Graduate School of Engineering, Osaka Metropolitan University, 1-1 Gakuen-cho, Naka-ku, Sakai, 599-8531, Japan*

[a)]Author to whom correspondence should be addressed: shikoh@omu.ac.jp (Eiji Shikoh)

ABSTRACT

Ferromagnetic resonance (FMR) excitation of a ferromagnetic metal single-layer film was tried by using a couple of permanent magnets as the source of the uniform static magnetic field and by using a co-planer waveguide connected with a network analyzer as the source of the radiofrequency (RF) magnetic field. A typical FMR spectrum of a ferromagnetic $Ni_{80}Fe_{20}$ thin film was successfully obtained and a clear electromotive force (EMF) was generated in the $Ni_{80}Fe_{20}$ thin film under the FMR excitation with the permanent magnets. That is, we achieved EMF generation in a $Ni_{80}Fe_{20}$ thin film under the FMR excitation with the permanent magnets, which can be applied as a simple energy transduction technology from ambient RF energy to electrical energy.

Development of energy harvesting technologies which harvest the existing micro-energy in an environment and transfer to electrical power is one important issue to efficiently use the Earth's natural resources.[1] In conventional, to use light, heat, vibration, electromagnetic field, and their related phenomena are focused as resources of the energy harvesting technologies.[1] Recently, by using an ambient radiofrequency (RF) energy at the frequency ($f$) of 2.4 GHz band, an energy harvesting with nanoscale spin rectifier (SR) arrays based on magnetic tunnel junctions (MTJs) has been achieved.[2] In the excellent work, the energy harvesting to the power of below -27 dBm has been achieved which is within the target power of the ambient RF energy from -82 to -20 dBm,[2] and the energy transduction efficiency was 10,000 mV/mW in cases with the single SR device based on MTJs to the RF power from -62 to -20 dBm, and 34,500 mV/mW in cases with the SR arrays based on MTJs to the RF power of -50 dBm.[2] This may be one successful example of the energy harvestings with spintronics.

Meanwhile, an electromotive force (EMF) generation phenomenon in a ferromagnetic metal (FM) thin film itself under the ferromagnetic resonance (FMR) has been discovered under the FMR excitation by using an electron spin resonance (ESR) system.[3,4] FMR is a magnetic dynamics phenomenon in a ferromagnetic material which is excited by using both a static magnetic field and an RF magnetic field in the GHz band, in general.[5] The generated EMF in FM films under the FMR excitation originates from various spin current rectification phenomena, such as the inverse spin-Hall effect,[6,7] the anomalous Hall effect,[8] the anisotropic magnetoresistance effect,[9] and so on. That is, if such EMFs are generated in an FM film under the FMR excited by using both permanent magnets as the static magnetic field source and an ambient RF energy in an environment as an RF magnetic field source, it can also be another energy transduction technology from ambient RF energy to electrical energy with spintronics. The FMR is excited with the FMR condition in an in-plane field according to the Kittel formula:[10]

$$\frac{\omega}{\gamma} = \sqrt{H_{FMR}(H_{FMR} + 4\pi M_S)}, \tag{1}$$

where $\omega$, $\gamma$, $H_{FMR}$ and $M_S$ are the angular frequency ($2\pi f$), the gyromagnetic ratio for FM film calculated from the Lande g-factor, the FMR field and the saturation magnetization of the FM film, respectively. The energy transduction technology with the FMR can correspond to multi-frequent RF energy, satisfied with the FMR condition,[10] which is a certain merit to use the FMR for the energy harvesting technology. An electrical charging experiment using the EMF generation phenomenon in an FM film under the FMR was also demonstrated by using an ESR system to easily excite the FMR,[11] although using an ESR system is out of the energy harvesting definition because large electrical power is necessary to operate a general ESR system.

In this study, the FMR excitation and EMF generation in an FM film under the FMR were challenged with permanent magnets, towards the development of a real energy harvesting technology with FMR. To the best of our knowledge, the FMR excitation with permanent magnets as the source of the static magnetic field has not been reported yet, while the nuclear magnetic resonance excitation, which is a kind of magnetic resonance phenomena, was achieved with permanent magnets on the development process of the table-top magnetic resonance imaging system.[12]

Figure 1(a) shows a schematic illustration of our sample and evaluation method. A $Ni_{80}Fe_{20}$ (Kojundo Chemical Laboratory Co., Ltd., 99.99% purity) was evaporated on a thermally oxidized silicon substrate using an electron beam deposition system at pressure <$10^{-6}$ Pa, and the deposition rate of 0.05 nm/s. The substrate temperature during $Ni_{80}Fe_{20}$ deposition was ambient (not controlled). No protection layer to prevent the $Ni_{80}Fe_{20}$ films from oxidizing was formed because only the FMR excitation and the EMF generation phenomenon in the $Ni_{80}Fe_{20}$ films under the FMR with permanent magnets were focused in this study, not the individual origins of EMFs. After the $Ni_{80}Fe_{20}$ deposition, the sample substrates were cut to the designed size as shown in Fig.

1(a).

A couple of cylindrical ferrite magnets to make the uniform static magnetic field ($H$) for FMR excitation was purchased. In the case that the two cylindrical magnets orient as shown in Fig. 1(b), the static magnetic field at the center position between the two cylindrical magnets $H(x)$ is calculated with the following equation under simple electromagnetics:

$$H(x) = H_0 \left\{ \frac{h+x}{\sqrt{r^2+(h+x)^2}} - \frac{x}{\sqrt{r^2+x^2}} \right\}, \qquad (2)$$

where $H_0$, $r$, and $h$ are the magnetic field at the top face, the radius, and the height of the cylindrical magnets, respectively. $x$ is the distance from the top face of one cylindrical magnet and equal to the center position between the two permanent magnets in this study. Permanent magnets with the $r$ of over 40 mm were selected to widely keep the uniform static magnetic field space. The $H$ applied with permanent magnets in this study was adjusted by moving the magnet positions and confirmed with a gaussmeter before and after the FMR experiments. The maximum $H$ applied with permanent magnets at the $x$ in our experimental setup was 1,030 Oe. To apply the RF magnetic field for FMR excitation, a coplanar waveguide connected to a vector network analyzer (VNA: KEYSIGHT Technology, N5221A) was located between the permanent magnets as shown in Fig. 1(a). Although utilizing ambient RF field is necessary for a real energy harvesting with FMR, it is one of future works. A nano-voltmeter (Keithley Instruments, 2182A) to detect electromotive forces from the samples was used. Leading wires for detecting the output voltage properties were directly attached at both ends of samples with silver paste. A thin insulating tape was stuck on the top of the coplanar waveguide to electrically separate from samples. The surface of $Ni_{80}Fe_{20}$ film samples is faced on the signal line of the coplanar waveguide. Figs. 1(c) & (d) show photographs of our experimental setup. While permanent magnets have not been placed yet in Fig. 1(c), our full setup for experiments with permanent magnets has prepared in Fig. 1(d). For a control experiment, an experimental setup with a couple of electromagnets instead of permanent magnets were also prepared.

First, to check the magnetic properties of our $Ni_{80}Fe_{20}$ films and the setup, electromotive force properties generated in a $Ni_{80}Fe_{20}$ film under the FMR were investigated by using the setup with a couple of electromagnets, not with permanent magnets. Figure 2(a) shows a typical FMR spectrum for a $Ni_{80}Fe_{20}$ film at the RF field frequency of 5 GHz. In Fig. 2(a), the longitudinal axis of the $\Delta S_{21}$ corresponds to the RF energy absorbance defined as the difference between the transmission properties of the VNA electrical power signals with and without the $Ni_{80}Fe_{20}$ film sample, and the electrical power of the input signal ($P$) from the VNA is +13 dBm (20 mW) which is the maximum $P$ of our setup. A clear absorbance signal has appeared around the $H$ of 300 Oe. Using the FMR condition of the eq. (1), the $M_S$ of the $Ni_{80}Fe_{20}$ film was evaluated to be 729 G which is comparable with previous reports.[3,6,13] Fig. 2(b) shows an EMF property generated in the same $Ni_{80}Fe_{20}$ film as Fig. 2(a) under the FMR, at the RF field frequency of 5 GHz and the $P$ of +13 dBm. A clear EMF signal is observed around 300 Oe corresponding to the $H_{FMR}$ of the $Ni_{80}Fe_{20}$ film and about 0.6 µV at the $P$ of +13 dBm has been obtained.

Next, the FMR excitation with permanent magnets was challenged. Figure 3(a) shows a typical RF field frequency-dependent FMR spectrum for a $Ni_{80}Fe_{20}$ film at the $H$ of 400 Oe applied by using permanent magnets. The longitudinal axis of the $\Delta S_{21}$ in Fig. 3(a) corresponds to the RF energy absorbance defined as the difference between the transmission properties of the VNA electrical power signals with and without the permanent magnets, and the $P$ is +13 dBm (20 mW). A clear absorbance signal has appeared at 6.3 GHz. Using the FMR condition of the eq. (1), the $M_S$ of our $Ni_{80}Fe_{20}$ film was evaluated to be 808 G which is comparable with previous reports.[3,6,13] Fig. 3(b) shows an RF field frequency-dependent EMF property generated under the FMR of the same $Ni_{80}Fe_{20}$ film as Fig. 3(a), at the $H$ of 400 Oe and the $P$ of +13 dBm. Each data in Fig. 3(b) was obtained at the frequency-fixed RF source and the *H* of 400 Oe. A clear EMF signal is observed at around 6 GHz corresponding to the FMR frequency of the $Ni_{80}Fe_{20}$ film and 0.2 μV at +13 dBm has been obtained. To make sure, a $Ni_{80}Fe_{20}$ film sample was newly prepared, and tested. As the result, as expected, the same characteristics as the Fig. 3 were observed. Moreover, to confirm that the above things come from the FMR excited with the permanent magnets, the *H* was changed by exchanging the permanent magnets with others making different *H* and with changing the *x*. In all control experiments, the FMR properties satisfied with the eq. (1) were observed. That is, the FMR excitation of a $Ni_{80}Fe_{20}$ film with permanent magnets was achieved and the EMF was successfully generated in the $Ni_{80}Fe_{20}$ film under the FMR excitation with permanent magnets.

Figure 4 shows the *P* dependence of the EMFs generated in another $Ni_{80}Fe_{20}$ film under the FMR excitation with permanent magnets, with the *H* of 415 Oe and the *f* of 6.13 GHz. The EMF is almost linearly increased with the *P*, similarly to the previous reports.[3,11] EMFs at below the *P* of 4 mW were under the measurement limit of our setup. The maximum energy transduction efficiency to electrical energy per our single $Ni_{80}Fe_{20}$ film of 10 nV/mW calculated from the data of the Fig. 3(b) is almost same as the cases with electromagnets for the static magnetic field. The energy transduction efficiency of our current setup is extremely small compared to the SR arrays demonstration,[2] although the energy transduction efficiency is easily improved with a series connection of lots of $Ni_{80}Fe_{20}$ film. At present, our result is only a demonstration of a basis of energy transduction technology with FMR excited by using permanent magnets, however, this is a certain step with realization of a simple energy harvesting technology with FMR because no electrical power is needed to apply the static magnetic field. If realized, it can harvest multi-frequency RF field energy more simply, satisfied with the eq. (1).[10] The simple energy transduction due to the FMR with ambient multi-frequent RF fields is demonstrated near future, which has a certain merit to directly use the FMR for the energy harvesting technology.

In summary, FMR excitation was achieved with permanent magnets, and EMF generation in a ferromagnetic $Ni_{80}Fe_{20}$ thin film under the FMR excitation by using permanent magnets was successfully demonstrated. While the energy transduction efficiency from the RF energy to electrical energy of 10 nV/mW is small, this achievement can be applied as a simple energy transduction technology from ambient RF energy to electrical energy.

This research was partly supported by a research grant from the Murata Science Foundation, a research grant from the Mazda Foundation, and a research grant from Iketani Science and Technology Foundation.

AUTHOR DECLARATIONS

Conflict of Interest

The authors declare no competing financial interests.

Author Contributions

Ryutoku Fujii: Conceptualization (equal); All experiments including setups; Data analysis (equal); Discussion (equal); Writing – original draft (equal).

Eiji Shikoh: Conceptualization (equal); Funding acquisition; Project administration; Supervision; Experimental system setup (partly support); Data analysis (equal); Discussion (equal); Writing – original draft (equal); Editing.

DATA AVAILABILITY

The data that supports the findings of this study are available from the corresponding authors upon reasonable request.

REFERENCES

[1] H. Akinaga, H. Fujita, M. Mizuguchi, and T. Mori, Sci. Technol. Adv. Mater. **19**, 543 (2018).

[2] R. Sharma, T. Ngo, E. Raimondo, A. Giordano, J. Igarashi, B. Jinnai, S. Zhao, J. Lei, Y-X Guo, G. Finocchio, S. Fukami, H. Ohno, and H. Yang, Nat. Electron. **7**, 653 (2024).

[3] A. Tsukahara, Y. Ando, Y. Kitamura, H. Emoto, E. Shikoh, M.P. Delmo, T. Shinjo, and M. Shiraishi, Phys. Rev. B **89**, 235317 (2014).

[4] K. Kanagawa, Y. Teki, and E. Shikoh, AIP Advances **8**, 055910 (2018).

[5] C. Kittel, Introduction to Solid State Physics (8$^{th}$ ed.), Wiley (2004).

[6] E. Saitoh, M. Ueda, H. Miyajima, and G. Tatara, Appl. Phys. Lett. **88**, 182509 (2006).

[7] S. O. Valenzuela, and M. Tinkham, Nature **442**, 176 (2006).

[8] R. Karplus, R. and J. M. Luttinger, Phys. Rev. 95, 1154 (1954).

[9] T. McGuire, and R. Potter, IEEE Trans. Magn. **11**, 1018 (1975).

[10] C. Kittel, Phys. Rev. **73**, 155 (1948).

[11] Y. Nogi, Y. Teki, and E. Shikoh, AIP Advances **11**, 085114 (2021).

[12] T. Oka, N. Inoue, S. Tsunoda, J. Ogawa, S. Fukui, K. Yokoyama, N. Sakai, M. Miryala, M. Murakami, M. Takahashi, and T Nakamura, Mater. Res. Express **7**, 056001 (2020).

[13] T. Seki, Y.-C. Lau, S. Iihama, and K. Takanashi, Phys. Rev. B **104**, 094430 (2021).

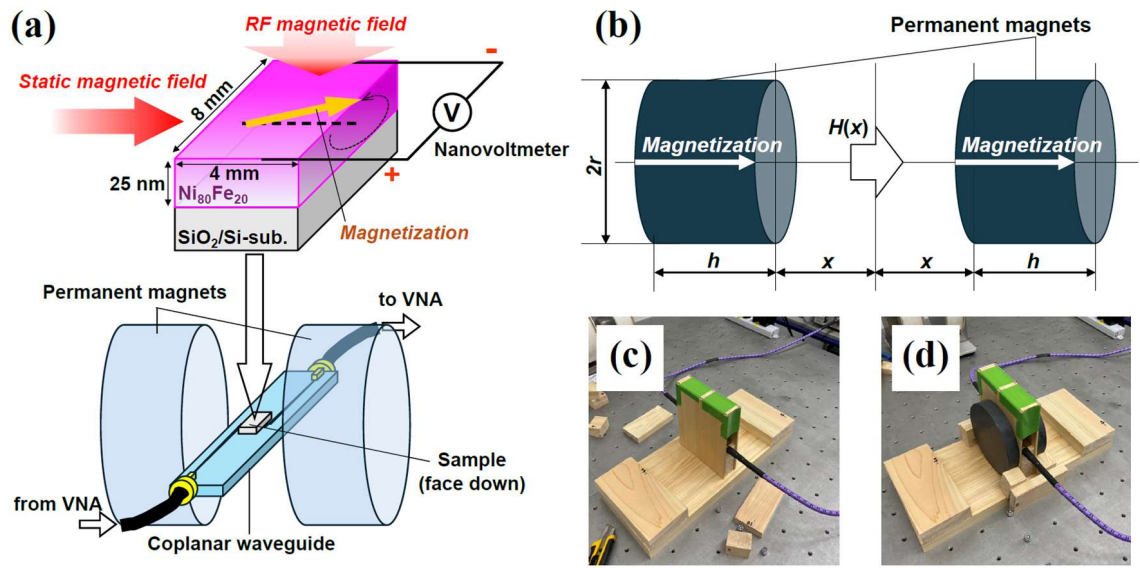

Fig. 1. (a) A schematic illustration of our sample structure and experimental setup. (b) A schematic illustration of the simple calculation model of the static magnetic field $H(x)$ at the center position between the two cylindrical magnets. Photographs of our experiment setup (c) without and (d) with permanent magnets for FMR excitations.

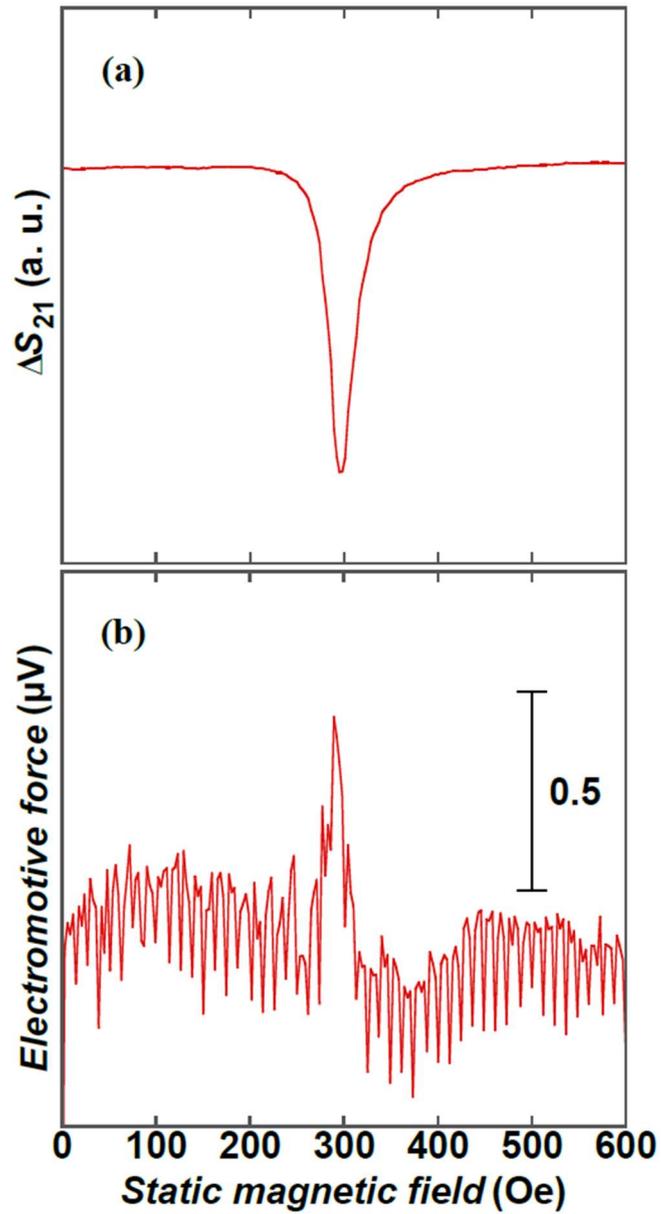

Fig. 2. (a) A typical FMR spectrum for a $Ni_{80}Fe_{20}$ film investigated with electromagnet, and (b) EMF properties generated in the same $Ni_{80}Fe_{20}$ film under the FMR excited by using a couple of electromagnets. The RF field frequency and power are 5 GHz and +13dBm (20 mW), respectively.

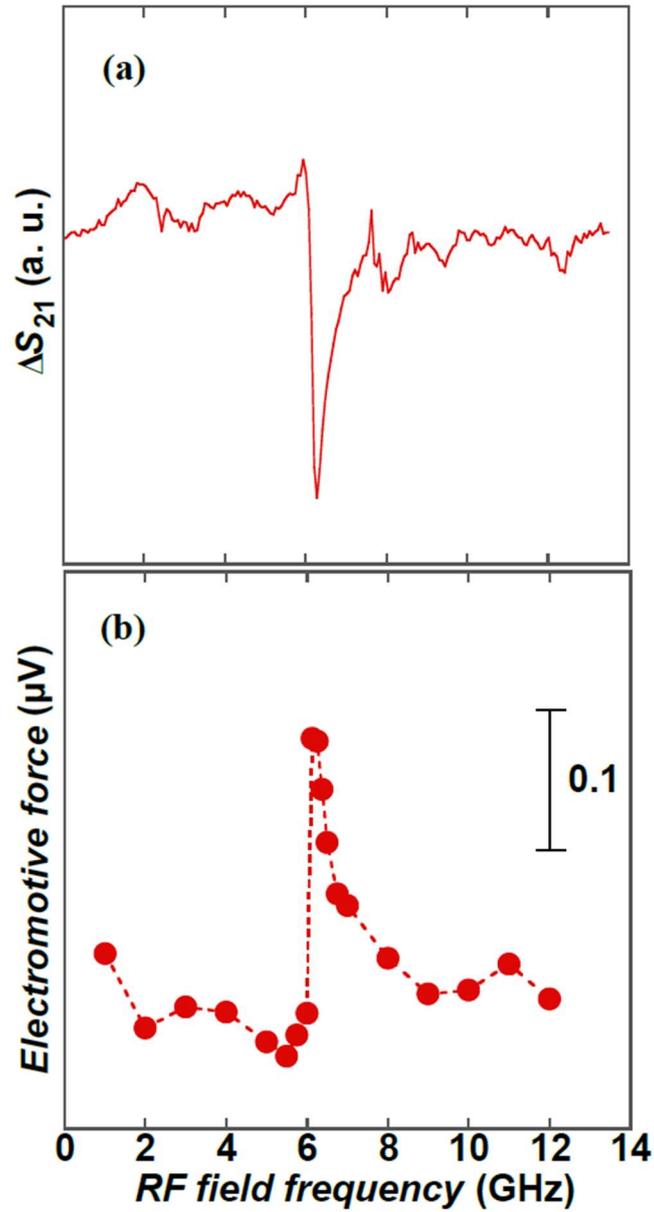

Fig. 3. (a) A typical RF field frequency-dependent FMR spectrum for a $Ni_{80}Fe_{20}$ film with permanent magnets, and (b) EMF properties generated in the same $Ni_{80}Fe_{20}$ film under the FMR excited by the permanent magnets. The static magnetic field at the $Ni_{80}Fe_{20}$ film sample position is 400 Oe and the RF power of +13dBm (20 mW).

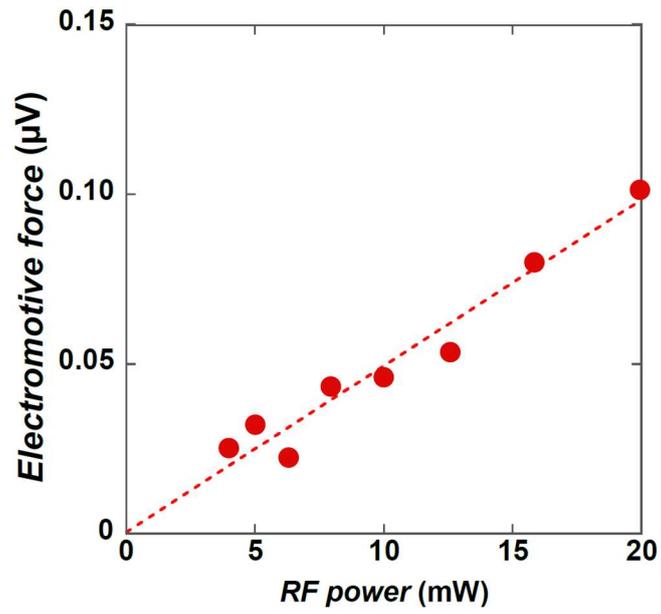

Fig. 4. An RF power dependence of the electromotive forces generated in a $Ni_{80}Fe_{20}$ film under the FMR excitation with permanent magnets. The static magnetic field at the $Ni_{80}Fe_{20}$ film sample position is 415 Oe and the RF field frequency is 6.13 GHz. The dash line is drawn just for guides.